\documentclass[12pt]{article}
\usepackage{graphicx}
\usepackage{amsmath}
\usepackage{amsfonts}
\usepackage{amssymb}    
\usepackage{graphicx}   
\usepackage{cite} 
\newcommand{\be}{\begin{equation}}
\newcommand{\ee}{\end{equation}}
\newcommand{\bea}{\begin{eqnarray}}
\newcommand{\eea}{\end{eqnarray}}
\newcommand{\beq}{\begin{equation}}
\newcommand{\eeq}{\end{equation}}
\newcommand{\ba}{\begin{array}}
\newcommand{\ea}{\end{array}}
\newcommand{\beqa}{\begin{eqnarray}}
\newcommand{\eeqa}{\end{eqnarray}}

\newcommand{\cL}{{\cal L}}

\newcommand{\cO}{{\cal O}}
\newcommand{\cB}{{\cal B}}

\newcommand{\no}{\nonumber}

\newcommand{\RE}{{\rm Re}}
\newcommand{\IM}{{\rm Im}}
\newcommand\mgl{{\tilde M}_g}

\def\kpn{K^+\rightarrow\pi^+\nu\bar\nu}
\def\klpn{K_{\rm L}\rightarrow\pi^0\nu\bar\nu}

\def \text{\mathrm}

\def\npb#1#2#3{    { Nucl. Phys. }{\bf B #1} (#2) #3}
\def\plb#1#2#3{    { Phys. Lett. }{\bf B #1} (#2) #3}

\def\prd#1#2#3{    { Phys. Rev. }{\bf D #1} (#2) #3}

\def\prl#1#2#3{    { Phys. Rev. Lett. }{\bf #1} (#2) #3}

\def\ijmpa#1#2#3{  { Int. J. Mod. Phys. }{\bf A #1} (#2) #3}

\def\jhep#1#2#3{   { JHEP  }{\bf #1} (#2) #3}

\begin{document}

\begin{flushright}
January 2006 \\
\end{flushright}
\vskip   1 true cm 
\begin{center}
{\Large \textbf{Higgs--mediated $K \rightarrow \pi \nu\overline{\nu}$ in 
the MSSM at large $\tan \beta$}}  \\ [20 pt]
\textsc{Gino Isidori}${}^{1}$ and \textsc{Paride Paradisi}${}^{2}$
  \\ [20 pt]
${}^{1}~$\textsl{INFN, Laboratori Nazionali di Frascati, I-00044 Frascati,
      Italy} \\ [5pt]
${}^{2}~$\textsl{ INFN, Sezione di Roma II and Dipartimento di Fisica, 
\\ Universit\`a di Roma
``Tor Vergata'', I-00133 Rome, Italy. }

\vskip   1 true cm 

\textbf{Abstract}
\end{center}
\noindent  
We analyze the impact of Higgs-mediated amplitudes on the rare decays 
$K_L\rightarrow\pi^0\nu\overline{\nu}$ and $K^{+}\rightarrow\pi^{+}\nu\overline{\nu}$
in the MSSM with large  $\tan \beta$ and general flavour mixing. 
We point out that, going beyond the minimal flavour violation hypothesis,  
$Z$-penguin amplitudes generated by  charged-Higgs exchange can induce 
sizable  modifications of $K \rightarrow \pi \nu\overline{\nu}$ rates. 
Interestingly, these effects scale as $\tan^4 \beta$ at the amplitude level.
For large values of   $\tan \beta$, this mechanism allows 
deviations from the SM expectations
even for tiny (CKM-type) off-diagonal mixing 
terms in the right-handed squark sector.

\vskip   1 true cm 

\paragraph{I.}
The strong suppression within the Standard Model (SM)
and the high sensitivity to physics beyond the SM
of the two $K \rightarrow \pi \nu\overline{\nu}$ rates,
signal the unique possibilities offered by these rare 
processes in probing the underlying mechanism of flavour mixing 
(see e.g.~Refs.~\cite{My,BSU}).
This statement has been reinforced by the 
recent theoretical progress in the evaluation 
of NNLO~\cite{NNLO} and power-suppressed~\cite{IMS}
contributions to $K^{+}\rightarrow\pi^{+}\nu\overline{\nu}$.
These allow to obtain predictions for the corresponding 
branching ratio of high accuracy ($\sim 6\%$), not far from 
exceptional level of precision ($\sim 2\%$) already reached 
in the $K_L\rightarrow\pi^0\nu\overline{\nu}$  case~\cite{BSU,BI}.

The non-standard contributions to $K \to \pi \nu\overline{\nu}$
decays appearing in the minimal supersymmetric
extensions of the SM (MSSM) with generic sources of flavour 
mixing have been widely discussed in the literature
\cite{Kpnn_SUSY1,CI,Jager,Kpnn_LNF}.
However, all previous analysis are focused on scenarios 
with $\tan\beta = O(1)$. In this case, taking into account 
the existing constraints from 
other flavour-changing neutral current (FCNC)
observables, the only amplitudes which allow sizable deviations from the SM  
are those generated by chargino/up-squark 
loops. In particular, large effects 
are generated only if the left-right mixing of  
the up squarks has a flavour structure substantially
different from the so-called Minimal Flavour 
Violation (MFV) hypothesis~\cite{MFV_SUSY}.

In this paper we point out the existence of a completely different
mechanism which could allows sizable 
modifications of $K \to \pi \nu\overline{\nu}$ amplitudes:
the charged-Higgs/top-quark loops in the large  $\tan\beta$
regime.\footnote{~As recently shown in Ref.~\cite{kl2},  
charged Higgs can induce another very interesting 
non-standard phenomenon in the kaon system:
violations of lepton universality which can reach the 
$1\%$ level in the 
$\Gamma(K\rightarrow e\nu)/\Gamma(K\rightarrow \mu\nu)$ ratio.} 
The anomalous coupling of charged-Higgs bosons 
responsible for this  effect is quite similar to 
the phenomenon of the Higgs-mediated FCNC currents which 
has been widely discussed in the literature 
(see e.g.~Refs.~\cite{Babu,others,Chanko,IR2,Foster}).
They are both induced by effective 
non-holomorphic terms and do not decouple 
in the limit of heavy squark masses 
(provided the Higgs sector remains light).
However, contrary to the neutral--Higgs 
contribution to $B_{s,d}\to \mu^+\mu^-$, the 
charged--Higgs contribution to 
$K \to \pi \nu\overline{\nu}$ can compete with the SM one 
only in presence of non-MFV soft-breaking terms.

The generic structure of the charged-Higgs couplings 
in the MSSM, at large $\tan\beta$, and with arbitrary 
flavour mixing in the squark sector, has been discussed 
first in Ref.~\cite{Foster}. There the phenomenological implications 
of these effects in rare $B$ decays, and the corresponding 
bounds on the off-diagonal soft-breaking terms have also 
been analyzed \cite{Foster}. As we will show, the  $B$-physics
constraints still allow a large room of  
non-standard effects in $K \to \pi \nu\overline{\nu}$:
in the kaon sector these Higgs-mediated amplitudes 
allow observable effects even for flavour-mixing terms of CKM size 
[in particular for $(M^{D}_{RR})_{ij} \sim (V_{CKM})_{ij} (M^{D}_{RR})_{jj}$].
Future precision measurements of $K \to \pi \nu\overline{\nu}$
rates could then provide the most stringent 
bounds, or the first evidence, of this type 
of effects.

\paragraph{II.}
The dimension-six effective Hamiltonian relevant for $\kpn$ 
and $\klpn$ decays in the general MSSM can be written as:
\be
\label{hMSSM} 
{\mathcal H}_{\rm eff}= \frac{G_{\rm F}}{\sqrt 2} \frac{\alpha}{2\pi}
\sin^2\theta_{\rm w} 
\left[ 
{\mathcal H}_{\rm eff}^{(c)}+{\mathcal H}_{\rm eff}^{(s.d.)}\right]
 ~+~ {\rm h.c.}
\ee
where ${\mathcal H}_{\rm eff}^{(c)}$ denotes the operators 
which encode physics below the electroweak 
scale (in particular charm quark loops) 
and are fully dominated by SM contributions \cite{NNLO}, 
while 
\be
\label{Ht}
{\mathcal H}_{\rm eff}^{(s.d.)}=
\sum_{l=e,\mu,\tau} V^{\ast}_{ts}V_{td} 
\left[X_L (\bar sd)_{V-A}(\bar\nu_l\nu_l)_{V-A}+ 
X_R (\bar sd)_{V+A}(\bar\nu_l\nu_l)_{V-A}\right]
\ee
denotes the part of the  effective Hamiltonian 
sensitive to short-distance dynamics.

Within the Standard Model ${X_R=0}$ and
$X_L$ is a real function (thanks to the normalization 
with the CKM factor 
$V^{\ast}_{ts}V_{td}\equiv \lambda_t$)~\cite{BSU,NNLO}:
\be
X_L^{\rm SM} = 1.464\pm 0.041~.
\label{eq:xsm}
\ee
Within the MSSM, in general both $X_R$ and  $X_L$ are not vanishing, 
and the misalignment between quark and squark flavour 
structures implies that are both complex quantities. 
Since the $K\to\pi$ matrix elements of $(\bar s d)_{V-A}$ and
$(\bar s d)_{V+A}$ are equal, the 
following combination
\be
X=X_L+X_R
\label{eq:Xtot}
\ee
allows us to describe all the short-distance 
contributions to $K \to \pi \nu\overline{\nu}$ decays.
In terms of this quantity the two branching ratios 
can be written as:
\bea
\label{bkpnZ}
\cB(\kpn) &=& \kappa_+\left[\left(\frac{\IM(\lambda_t X)}{\lambda^5}\right)^2 
+ \left(  P_{(u,c)} + \frac{\RE(\lambda_t X)}{\lambda^5}\right)^2\right]~, \\
\label{bklpnZ}
\cB(\klpn) &=& \kappa_L \left( \frac{\IM(\lambda_t X)}{\lambda^5}\right)^2~,
\eea
where $\lambda = |V_{us}| = 0.225$, the $\kappa$-factors are
$\kappa_+ = (4.84\pm 0.06) \times  10^{-11}$
and  $\kappa_L = (2.12\pm 0.03) \times 10^{-10}$~\cite{BSU},
and the contribution arising from charm and light-quark 
loops is $P_{(u,c)} = 0.41\pm 0.04$~\cite{NNLO,IMS}.

\paragraph{III.}
Within a two-Higgs model of type-II (such as the MSSM
Higgs sector, if we neglect the non-holomorphic terms
which arise beyond the tree level \cite{HRS}),
the $Z$-penguin amplitudes generated 
by charged-Higgs/top-quark diagrams (at the one loop level) 
give:
\be
X^{H^{\pm} }_L  = - \frac{m_t^2 \cot^2\beta}{2 M_W^2 } f_H (y_{tH})~,
\qquad 
X^{H^{\pm} }_R =
\frac{m_{d}m_{s}\tan^2\beta}{2M_W^2} f_H (y_{tH})~,
\ee
where $y_{tH}=m_t^2/M_H^2$ and 
\be
f_H (x) = \frac{x}{4(1-x)}+\frac{x\log x}{4(x-1)^2}~.
\label{floop}
\ee
The numerical impact of these contributions is quite small
\cite{Kpnn_SUSY1,CI,Jager}:
the lower bounds on $M_H$ (the charged-Higgs mass)
and $\tan\beta$ imply that $X^{H^{\pm} }_L$ 
can at most reach few \% compared to $X^{\rm SM}_L$;
the possible  $\tan\beta$ enhancement of 
$X^{H^{\pm} }_R$ is more than compensated 
by the smallness of $m_{d,s}$.
The basic feature of the interesting new 
phenomenon appearing beyond the one-loop level 
(with generic flavour mixing and large $\tan\beta$) 
is a contribution to $X_R$ not suppressed by $m_{d,s}$.

Following the procedure of Ref.~\cite{IR2,Foster},
in order to compute all the non-decoupling effects 
at large $\tan\beta$ one needs to:
i) evaluate the effective dimension-four operators 
appearing at the one-loop level which 
modify the tree-level Yukawa Lagrangian;
ii) expand the off-diagonal mass terms in the squark 
sector by means of the mass-insertion approximation \cite{GGMS};
iii)  diagonalize the quark mass terms and derive 
the effective interactions between quarks and 
heavy Higgs fields. In the $\bar{U}^i_L D^j_R H^+$ 
case, this leads to \cite{Foster}
\be
\cL_{H^+}= \frac{g }{\sqrt{2}M_W} \frac{  \tan\beta  }{ (1+\epsilon_i \tan\beta) }
\left[ m_{d_j}  V_{ij} +  m_{d_i} (\Delta_{RR})_{ij}
+ {\tilde M}_g (\Delta_{LR})_{ij}
 \right]~\bar{U}^i_L D^j_R H^+ ~+~ {\rm h.c.}~,
\label{eq:effH}
\ee
where 
\be
 (\Delta_{RR})_{ij} = 
\frac{  \tan\beta \left(\delta^d_{RR}\right)_{ij}}{(1+\epsilon_i \tan\beta) }
~ \epsilon_{RR} ~, \qquad 
(\Delta_{LR})_{ij} = \left(\delta^d_{RL}\right)_{ji}
~ \epsilon_{i}~\epsilon_{RL}~,
\ee
and the $\delta$'s are defined as in \cite{Chanko}.
Following standard notations, we have used
\bea
&& \epsilon_{i} = - \frac{2\alpha_s}{3\pi}\frac{\mu}{\mgl} H_2\left(x_{d_{_R} g},x_{d_{_L} g}\right)
-\delta_{i3} \frac{ y_t^2 A_t}{16\pi^2\mu} H_2( x_{u_{_R} \mu}, x_{u_{_L} \mu})~, 
 \no\\
&& \epsilon_{RR} = \frac{2\alpha_s}{3\pi}\frac{\mu}{\mgl}x_{d_{_R}g}
H_3\left(x_{d_{_L}g},x_{d_{_L}g},x_{d_{_R}g}\right)~, \qquad 
\epsilon_{LR} = \frac{2\alpha_s}{3\pi} x_{d_{_{RL}}g}  
H_2\left(x_{d_{_R} g},x_{d_{_L} g}\right)~ \qquad
\eea
with $x_{q_{_K} g}=\tilde M_{q_{_K}}^2/\tilde M^{2}_{g}$
and $x_{q_{_{KS}} g}= \sqrt{\tilde M_{q_{_K}}^2 \tilde M_{q_{_S}}^2}/\tilde M^{2}_{g}$.
The explicit expressions of the $H_i$ one-loop functions, normalized by 
$H_2(1,1)=-1/2$ and $H_3(1,1,1)=1/6$, can be found 
for instance in Ref.~\cite{Foster}.

\begin{figure}[t]
\begin{center}
\includegraphics[scale=0.80]{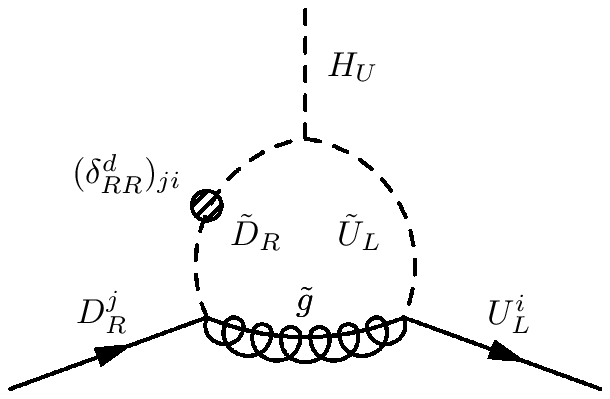} 
\hskip 0.8 cm 
\includegraphics[scale=0.80]{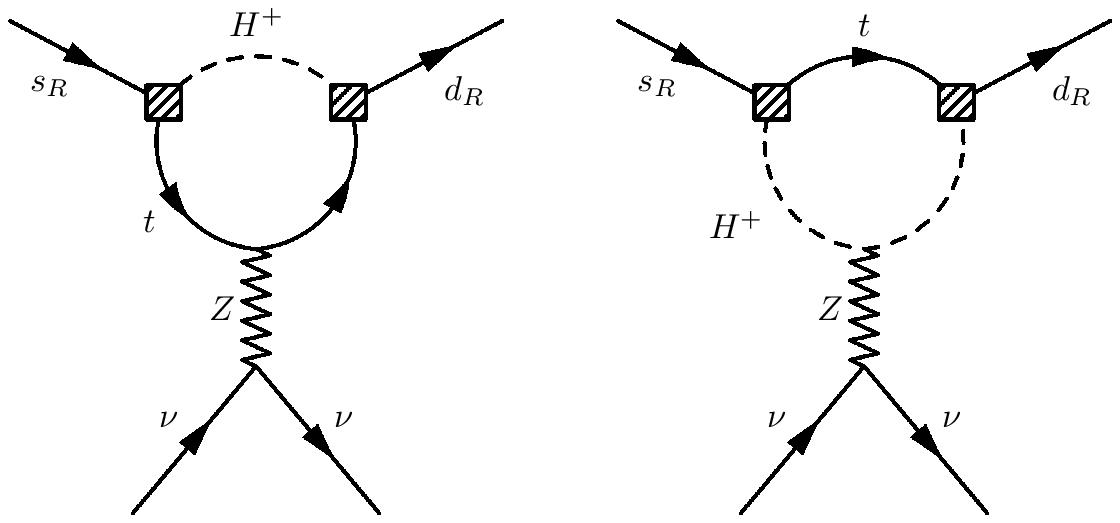}  
\caption{\label{fig} Left: leading contribution to the 
effective  $\bar{U}^i_L D^j_R H^+$ coupling in Eq.~(\ref{eq:effH}).
Right:~irreducible one-loop diagrams contributing to the 
$K\to \pi\nu\bar\nu$ amplitude in Eq.~(\ref{eq:XR_eff}). }
\end{center}
\end{figure}

As anticipated, the effective Lagrangian (\ref{eq:effH}) contains 
effective couplings not suppressed by the factor $m_{d_j}$
(which in our case corresponds to $m_{d}$ or $m_{s}$).
Since vacuum-stability arguments force 
$\left(\delta^d_{RL}\right)_{ji}$ to be proportional to
$m_{d_j}$ \cite{CaDimo}, only the right-right (RR) mixing terms
can effectively overcome this strong suppression. 
Evaluating the charged-Higgs/top-quark contributions 
to the $K\to\pi\nu\bar\nu$ amplitude using this effective 
Lagrangian (see Fig.~\ref{fig}), and retaining only the leading terms, we find 
\beqa
\left(X^{H^{\pm}}_{\rm eff}\right)_R
&=&\left[ \frac{(\delta^d_{RR})_{31} (\delta^d_{RR})_{32} }{\lambda_t}\right]
\left(\frac{m^{2}_{b}\tan^2\beta}{2 M_W^2}\right)
\frac{\epsilon^{2}_{RR}\tan^2\beta}{(1+\epsilon_{i}\tan\beta)^4}\,\,f_H(y_{tH})
\nonumber \\
&&+\left[ \frac{(\delta^d_{RR})_{31} }{ V_{td} }\right]
\left(\frac{m_{s}m_{b}\tan^2\beta}{2M_W^2}\right)
\frac{\epsilon_{RR}\tan\beta}{(1+\epsilon_{i}\tan\beta)^3}\,\,f_H(y_{tH})~, \qquad 
\label{eq:XR_eff}
\eeqa
which is the main result of the present paper. 

As can be noted, the first term on the r.h.s.~of Eq.~(\ref{eq:XR_eff}), 
which would appear only at the three-loop level in a {\em standard} loop 
expansion\footnote{~We denote by {\em standard} loop expansion the one 
performed without diagonalizing the effective one-loop Yukawa
interaction: as shown in Ref.~\cite{Babu}, the diagonalization of the quark 
mass terms  is a key ingredient to re-sum to all orders the large $\tan\beta$ corrections.} 
can be largely enhanced by the $\tan^4\beta$ factor
and does not contain any suppression due to light quark masses.
Similarly to the double mass-insertion mechanism of Ref.~\cite{CI},
also in this case the potentially leading effect is the one generated
when two off-diagonal squark mixing terms replace the two CKM 
factors $V_{ts}$ and $V_{td}$. For completeness, we have reported
in Eq.~(\ref{eq:XR_eff}) also the subleading term with a single
mass insertion, where only  $V_{td}$ is replaced by
a $3$--$1$ mixing in the squark sector.

\begin{table}[t]
\hskip -0.6 true cm 
\begin{tabular}{|c|c|c||c|c|c|c|}
\hline 
   $M_H$ 
 & \raisebox{-7pt}[0pt][0pt]{$x_{dg} $}
 & $\tilde M_{q}$
 & $\epsilon_K$ 
 & \raisebox{0pt}[12pt][0pt]{$K_L\rightarrow \mu^+\mu^-$} 
 & \raisebox{0pt}[12pt][0pt]{$K^+\rightarrow \pi^+\nu\bar{\nu}$} 
 & \raisebox{0pt}[12pt][0pt]{$K_L \rightarrow \pi^0 \nu\bar{\nu}$}  \\
\cline{4-7}  
   \raisebox{4pt}[14pt][0pt]{(GeV)}
 &
 & \raisebox{4pt}[15pt][0pt]{(GeV)}
 & \raisebox{2pt}[15pt][0pt]{$\left|\IM(\delta^{d}_{RR})_{21}\right|$}  
 & \raisebox{2pt}[15pt][0pt]{$\left|\RE(\delta^{d}_{RR})_{21}\right|$}   
 & \raisebox{2pt}[15pt][0pt]{$\left|\RE(\delta^{d}_{RR})_{23}(\delta^{d}_{RR})_{31}\right|$} 
 & \raisebox{2pt}[15pt][0pt]{$\left|\IM(\delta^{d}_{RR})_{23}(\delta^{d}_{RR})_{31}\right|$}   \\
\hline\hline
\cline{2-7}
    \raisebox{-7pt}[0pt][0pt]{ 500 } 
  & \raisebox{-7pt}[0pt][0pt]{  1  } 
  & 500  
  & $1.2\times 10^{-3}$
  & $3.1\times 10^{-2}$
  & $3.3\times 10^{-2}$
  & $8.7\times 10^{-3}$\\
\cline{3-4}
  &
  & 1000  & $2.6\times 10^{-3}$
  & ($2.2\times 10^{-3}$)
  & ($1.6\times 10^{-4}$)	
  & ($4.3\times 10^{-5}$) \\
\cline{2-7}
\hline
    \raisebox{-7pt}[0pt][0pt]{ 1000 } 
  & \raisebox{-7pt}[0pt][0pt]{  1  } 
  & 1500  
  & $4.0\times 10^{-3}$ 	
  & $1.3\times 10^{-1}$ 	
  & $7.8\times 10^{-2}$ 	
  & $2.1\times 10^{-2}$\\
\cline{3-4}
  &     
  & 2000  
  & $5.4\times 10^{-3}$ 	
  & ($8.8\times 10^{-3}$) 
  & ($3.9\times 10^{-4}$) 
  & ($1.0\times 10^{-4}$) \\ 
\hline
\end{tabular}
\caption{Comparison of the sensitivity to RR flavour-mixing couplings of different 
kaon observables. The
values in the two $K\rightarrow \pi\nu\bar{\nu}$ columns 
correspond to 
$10\%$ deviations of the corresponding rates compared to the SM.
For comparison, we report the bounds derived from the gluino/squark box 
amplitude in $\epsilon_K$ \cite{Roma} (under the assumption $\RE(\delta^{d}_{RR})_{21}=0$
and allowing up to 20\% non-standard contributions to $\epsilon_K$)  
and the neutral-Higgs penguin contribution to $K_L\rightarrow \mu^+\mu^-$ \cite{IR2}. 
All results are obtained for $\tan\beta=50$
and $|\mu|={\tilde M}_d$.  In the last three columns
the value outside (between) brackets correspond to $\mu>0$ ($\mu<0$).
\label{tab1}}
\end{table}

\begin{table}[t]
\begin{center}
\begin{tabular}{|c|c|c||c|c|c|c|}
\hline 
   $M_H$
 & \raisebox{-7pt}[0pt][0pt]{$x_{dg} $} 
 & $m_{\tilde{q}}$
 & \raisebox{0pt}[12pt][0pt]{$B_d$--$\bar B_d$} 
 & \raisebox{0pt}[12pt][0pt]{$B_{s,d} \rightarrow \mu^+\mu^-$} 
 & \multicolumn{2}{|c|}{\raisebox{0pt}[12pt][0pt]{$K_L \rightarrow \pi^0 \nu\bar{\nu}$} }\\
\cline{4-7}
   \raisebox{4pt}[14pt][0pt]{(GeV)}
 &
 & \raisebox{4pt}[15pt][0pt]{(GeV)}
 & \raisebox{2pt}[15pt][0pt]{$\left|\IM(\delta^{d}_{RR})_{31}\right|$}
 & \raisebox{2pt}[15pt][0pt]{$\left|(\delta^{d}_{RR})_{32,31}\right|$}
 & \raisebox{2pt}[15pt][0pt]{$\left|\IM(\delta^{d}_{RR})_{31}\right|$}
 & \raisebox{2pt}[15pt][0pt]{$\left|\IM(\delta^{d}_{RR})_{23}(\delta^{d}_{RR})_{31}\right|$}   \\
\hline\hline
    \raisebox{-7pt}[0pt][0pt]{ 500 }
  & \raisebox{-7pt}[0pt][0pt]{ 1 }
  &  500
  &  $1.0\times 10^{-1}$
  &  $0.1,0.065$
  &  $0.64$
  &  $8.7\times 10^{-3}$  \\
\cline{3-4}
  &
  & 1000
  &  $2.0\times 10^{-1}$
  &  $(0.01,0.007)$
  &  $(1.2\times 10^{-2})$
  &  $(4.3\times 10^{-5})$  \\
\hline\hline
    \raisebox{-7pt}[0pt][0pt]{ 1000 }
  & \raisebox{-7pt}[0pt][0pt]{ 1 }
  &  1500
  &  $3.0\times 10^{-1}$
  &  $0.4,0.26$
  &  $-$	
  &  $2.0\times 10^{-2}$  \\
\cline{3-4}
  &
  & 2000
  &  $4.0\times 10^{-1}$
  &  $(0.045,0.029)$
  &  $(2.8\times 10^{-2})$
  &  $(1.0\times 10^{-4})$  \\
\hline
\end{tabular}
\end{center}
\caption{Comparison of the sensitivity to RR flavour-mixing couplings of
 $K_L\rightarrow \pi^0\nu\bar{\nu}$ vs.~various $B$ physics observables.
The values reported in the two
$K_L \rightarrow \pi^0 \nu\bar{\nu}$ columns
correspond to $10\%$ deviations with respect to the SM,
for the leading (double MIA) and subleading (single MIA)
terms in Eq.~(\ref{eq:XR_eff}).
For comparison, we report the bounds derived from the gluino/squark box 
amplitude in $B_d$--$\bar B_d$ mixing \cite{Roma2} (taking into account the
interference with the SM contribution) 
and the neutral-Higgs penguin contribution to $B_s \rightarrow \mu^+\mu^-$ \cite{IR2}
(taking into account the recent experimental bound in \cite{Bmm}). 
As in Table~\ref{tab1},  all results are obtained for $\tan\beta=50$
and $\mu={\tilde M}_d$ ($\mu=-{\tilde M}_d$).
\label{tab2}}
\end{table}

\begin{figure}[t]
\begin{tabular}{cc}
\includegraphics[scale=0.4]{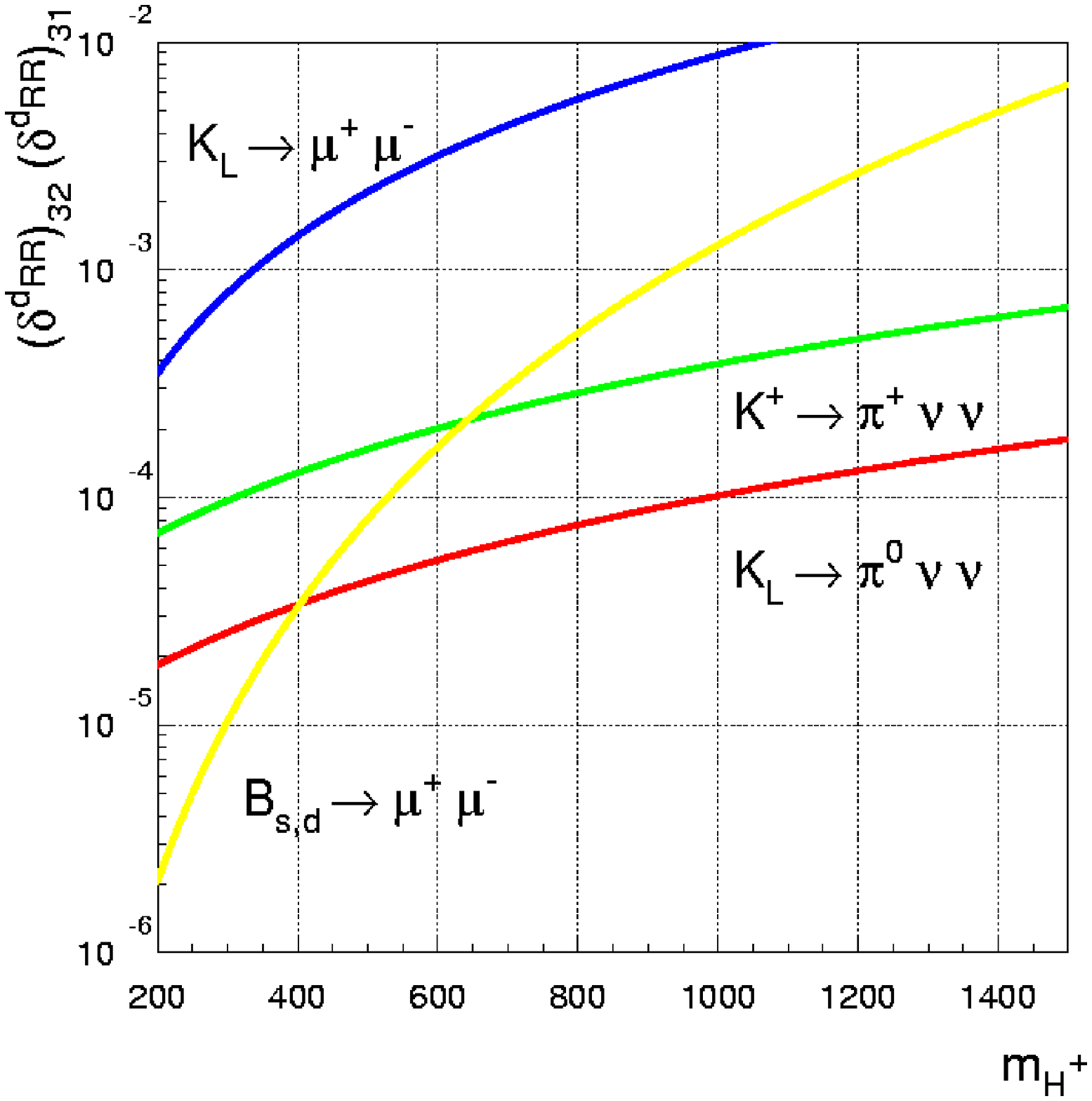} &
\includegraphics[scale=0.4]{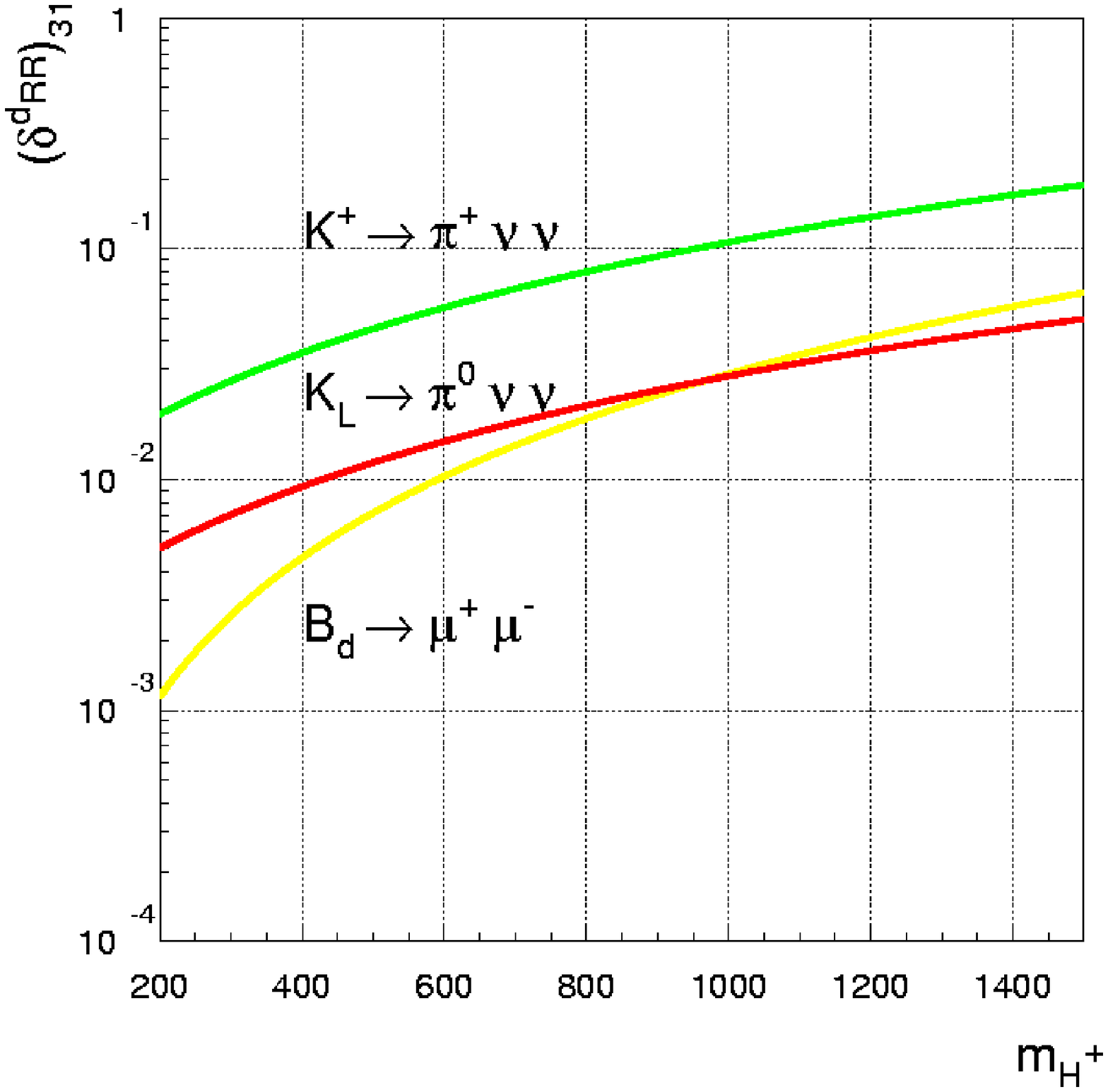}
\end{tabular}
\caption{\label{fig2}
Sensitivity to $(\delta^{d}_{RR})_{23}(\delta^{d}_{RR})_{31}$ and $(\delta^{d}_{RR})_{31}$
of various rare $K$ and $B$ decays as a function of the charged-Higgs boson mass. 
The bounds from the two $K\rightarrow \pi\nu\bar{\nu}$ modes are 
obtained under the assumption of a $10\%$ measurements of their branching ratios, 
and correspond to bounds on the real ($K^+ \to \pi^+\nu\bar\nu$) 
and the imaginary parts ($K_L \to \pi^0 \nu\bar\nu$) of the SUSY 
couplings. The $B_{s,d} \rightarrow \mu^+\mu^-$ bounds refer to the 
absolute value of the SUSY couplings and are based on the latest 
experimental limits \cite{Bmm}. All plots have been obtained with 
$\tan\beta\!=\!50$, $|\mu|={\tilde M}_d$, $\mu\!<\!0$, and $x_{dg}\!=\!1$.}
\end{figure}

We stress that the result in Eq.~(\ref{eq:XR_eff}) is the 
only potentially relevant Higgs-mediated contribution 
to $K\to\pi\nu\bar\nu$ amplitudes in the large $\tan\beta$
regime. Indeed, charged-Higgs contributions to $X_L$ do not receive 
any $\tan\beta$ enhancement even beyond the MFV hypothesis.
In principle, there are  $\tan^4\beta$ contributions 
to both  $X_L$ and  $X_R$  induced by neutral-Higgs exchange;
however, the presence of $b$ quarks inside the loops implies 
an additional suppression factor $\sim (m_b/m_t)^2$ 
of these contributions with respect to the 
charged-Higgs one in Eq.~(\ref{eq:XR_eff}).

\paragraph{IV.} The sensitivity of the two $K\rightarrow \pi\nu\bar{\nu}$
modes to the squark mixing terms appearing in  Eq.~(\ref{eq:XR_eff})
is illustrated in Tables~\ref{tab1}--\ref{tab2} and in 
Figure~\ref{fig2}. On general grounds, we can say that 
if $\tan\beta$ is large and  $\mu< 0$, the existing (and 
near-future) constraints on $(\delta^{d}_{RR})_{31}$ and $(\delta^{d}_{RR})_{32}$ 
from other $K$-- and $B$--physics observables still allow 
sizable non-standard contributions to the two 
$K\rightarrow \pi\nu\bar{\nu}$ rates.

The enhanced sensitivity of $K\rightarrow \pi\nu\bar{\nu}$ amplitudes 
to the $\mu<0$ scenario --at large  $\tan\beta$-- is due to the 
$(1+\epsilon_i\tan\beta)^{-n}$ factors (with $n=3,4$)
appearing in Eq.~(\ref{eq:XR_eff}).
For $\tan\beta \approx 50$ and degenerate SUSY particles,
one gets $(1+\epsilon_i\tan\beta)\approx (1+0.6~{\rm sgn}~\mu)$. 
Thus the bounds on $(\delta^{d}_{RR})_{32}(\delta^{d}_{RR})_{31}$
vary by a factor of $(1+0.6)^{4}/(1-0.6)^{4}\approx 250$ 
passing from $\mu< 0$ to $\mu> 0$. 
We stress that at the moment 
we don't have any clear phenomenological indication about the 
sign of $\mu$. A non-trivial constraint can be extracted 
from $\Gamma(B \to X_s \gamma)$:  charged-Higgs and chargino 
contributions to this observable must have opposite sign
in order to fulfill its precise experimental determination.
The relative sign between these two amplitudes is 
proportional to $ {\rm sgn}(\mu A_t)$,  
where $A_t$ is the trilinear soft-breaking term   
appearing in the stop mass matrix. Thus both signs 
of $\mu$ are allowed by $B \to X_s \gamma$, provided the sign 
of $A_t$ is chosen accordingly. In principle, an indication 
about the sign of $\mu$ could be deduced from the supersymmetric 
contribution to $(g_{\mu}-2)$. However, this indication is 
not reliable given the large theoretical uncertainties which 
still affect the estimate of the long-distance contributions
to this observable.

In Tables~\ref{tab1}--\ref{tab2} and in 
Figure~\ref{fig2} we have not explicitly reported bounds on  $(\delta^{d}_{RR})_{32}$
from $B \to X_s \gamma$ since the latter are very weak, both
for low $\tan\beta$ values (where the dominant contribution arises 
from gluino/squark penguins) and for large $\tan\beta$ values
(where the Higgs-mediated FCNC effects become sizable).
The weakness of the $B \to X_s \gamma$ bounds is a consequence of 
the absence of interference between
SM and SUSY contributions, when the latter are induced by down-type 
right-right flavour-mixing terms. This should be
contrasted with the $K \rightarrow \pi \nu\overline{\nu}$ case,
where SM and Higgs-mediated SUSY contributions have a maximal interference
[see Eq.~(\ref{eq:Xtot})].  Remarkably, this allows to 
generate $\cO(10\%)$ deviations from the SM expectations 
even for $(\delta^{d}_{RR})_{31} (\delta^{d}_{RR})_{32} = \cO(\lambda^5)$,
which can be considered as a natural reference value 
for possible violations quark/squark 
flavour alignment.

Another interesting feature of the supersymmetric charged-Higgs 
contribution to $K\rightarrow \pi\nu\bar{\nu}$ is 
the slow decoupling for $M_{H} \to \infty$: being generated by 
effective one-loop diagrams, the $X_R$ function develops 
a large logarithm and behaves as $X_R\sim y_{tH} \log(y_{tH})$
for $y_{tH}\!=\!m_t^2/M_H^2\!\ll 1$. 
On the contrary, the neutral-Higgs contribution to  
$B_{s,d}\rightarrow\mu^{+}\mu^{-}$ (and $K_{L}\rightarrow\mu^{+}\mu^{-}$),
which is generated by an effective tree-level diagram, 
does not contain any large logarithm. This is the reason why 
the $K\rightarrow \pi\nu\bar{\nu}$ constraints are much more effective 
than the $B_{s,d}\rightarrow\mu^{+}\mu^{-}$ ones for large 
$M_{H}$, as shown in Fig~\ref{fig2}.

\paragraph{V.} In summary, we have pointed out the
existence of a new mechanism which can induce
sizable non-standard effects in $\cB(K \to \pi \nu\overline{\nu})$ 
within supersymmetric extensions of the SM:
an effective FCNC operator of the type 
$\bar s_R \gamma^\mu d_R ~\bar\nu_L \gamma_\mu \nu_L$, generated by 
charged-Higgs/top-quark loops.
The coupling of this operator is phenomenologically relevant only at 
large $\tan\beta$ and with non-MFV right-right
soft-breaking terms: a specific but well-motivated scenario
within grand-unified theories (see e.g.~\cite{GUT}).
As we have shown, these non-standard effects do not 
vanish in the limit of heavy squarks and gauginos, and have 
a slow decoupling with respect to the charged-Higgs boson mass.
This implies that precision measurements of the two  $K\rightarrow \pi\nu\bar{\nu}$
rates, which should be within the reach of a new generation 
of dedicated experiments \cite{new_Kpnn}, could provide the 
most stringent bounds, or the first evidence, of this 
non-standard scenario.

\section*{Acknowledgments} 
This work is partially supported by IHP-RTN, 
EC contract No.\ HPRN-CT-2002-00311 (EURIDICE).


\begin{thebibliography}{999}
\footnotesize{

\bibitem{My}
G.~Isidori, \ijmpa {17}{2002}{3078} [hep-ph/0110255]; 
Annales Henri Poincare 4 (2003) S97 [hep-ph/0301159]; 
T.~Hurth, hep-ph/0511280.

\bibitem{BSU}
A.~J.~Buras, F.~Schwab and S.~Uhlig, hep-ph/0405132; 
D.~Bryman, A.~J.~Buras, G.~Isidori, L.~Littenberg, hep-ph/0505171.

\bibitem{NNLO}
A.~J.~Buras, M.~Gorbahn, U.~Haisch and U.~Nierste, hep-ph/0508165.

\bibitem{IMS}
G.~Isidori, F.~Mescia and C.~Smith,
Nucl.\ Phys.\ B {\bf 718} (2005) 319 [hep-ph/0503107].

\bibitem{BI}
G.~Buchalla and G.~Isidori,
Phys.\ Lett.~{\bf B440} (1998) 170 [hep-ph/9806501].

   
\bibitem{Kpnn_SUSY1}
Y.~Nir and M.P.~Worah,
\plb{423}{1998}{319} [hep-ph/9711215];
A.J.~Buras, A.~Romanino and L.~Silvestrini,
\npb{520}{1998}{3} [hep-ph/9712398];
A.J.~Buras {\em et al.},
\npb{566}{2000}{3} [hep-ph/9908371].

\bibitem{CI}
G. Colangelo and G. Isidori, JHEP {\bf 09} (1998) 009.

\bibitem{Jager} 
A.~J.~Buras, T.~Ewerth, S.~J\"ager and J.~Rosiek,
hep-ph/0408142.

\bibitem{Kpnn_LNF}
G.~Isidori, F.~Mescia, P.~Paradisi, C.~Smith, S.~Trine, 
in preparation (see talk by G.~I. at {\em Flavour in the
era of the LHC}, CERN, November 2005).

\bibitem{MFV_SUSY}
L.~J.~Hall and L.~Randall,
Phys.\ Rev.\ Lett.\  {\bf 65} (1990) 2939; 
G.~D'Ambrosio, G.~F.~Giudice, G.~Isidori and A.~Strumia,
{Nucl.\ Phys.}~{\bf B645} (2002) 155.

\bibitem{kl2}
A.~Masiero, P.~Paradisi and R.~Petronzio,
hep-ph/0511289.

\bibitem{Babu}
K.~S.~Babu and C.~Kolda,
\prl{84}{2000}{228} [hep-ph/9909476];
G.~Isidori and A.~Retico;
\jhep{0111}{2001}{001} [hep-ph/0110121].
A.~Dedes and A.~Pilaftsis,
\prd{67}{2003}{015012} [hep-ph/0209036].

\bibitem{others}
C.~Hamzaoui, M.~Pospelov and M.~Toharia,
\prd{59}{1999}{095005} [hep-ph/9807350];
C.~S.~Huang, W.~Liao and Q.~S.~Yan,
\prd{59}{1999}{011701} [hep-ph/9803460]; 
C.~Bobeth, T.~Ewerth, F.~Kruger and J.~Urban,
\prd{64}{2001}{074014} [hep-ph/0104284];
A.~Dedes, H.K.~Dreiner and U.~Nierste,
\prl{87}{2001}{251804} [hep-ph/0108037].

\bibitem{Chanko}
P.H.~Chankowski and L.~Slawianowska,
\prd{63}{2001}{054012} [hep-ph/0008046];
C.~Bobeth, T.~Ewerth, F.~Kruger and J.~Urban, 
\prd{66}{2002}{074021}, [hep-ph/0204225];
J.~Foster, K.~Okumura and L.~Roszkowski,
Phys.\ Lett.\ B {\bf 609} (2005) 102 [hep-ph/0410323].


\bibitem{IR2}
G.~Isidori and A.~Retico, \jhep{09}{2002}{063} [hep-ph/0208159].

\bibitem{Foster}
J.~Foster, K.~Okumura and L.~Roszkowski,
JHEP {\bf 0508} (2005) 094 [hep-ph/0506146].

\bibitem{HRS}
L.J.~Hall, R.~Rattazzi and U.~Sarid,
\prd{50}{1994}{7048} [hep-ph/9306309].

\bibitem{GGMS}
 F. Gabbiani, E. Gabrielli, A. Masiero and L. Silvestrini, \npb{477}{1996}{321}
 [hep-ph/9604387].


\bibitem{CaDimo}
J.A. Casas and S. Dimopoulos, 
Phys.\ Lett.\ B {\bf 387} (1996) 107
[hep-ph/9606237].  


\bibitem{Roma}    
M.~Ciuchini {\it et al.},
JHEP {\bf 9810}, 008 (1998)
[hep-ph/9808328].

\bibitem{Roma2}
D.~Becirevic {\it et al.},
Nucl.\ Phys.\ B {\bf 634} (2002) 105 [hep-ph/0112303].


\bibitem{Bmm}
R.~Bernhard {\it et al.}  [CDF Collaboration],
hep-ex/0508058.


\bibitem{GUT}
T.~Moroi, \plb{493}{2000}{366}
[hep-ph/0007328];
D.~Chang, A.~Masiero and H.~Murayama,
Phys.\ Rev.\ D {\bf 67} (2003) 075013
[hep-ph/0205111].

\bibitem{new_Kpnn}
G.~Anelli {\it et al.}
{\em Proposal to measure the rare decay $K^+ \to \pi^+ \nu \bar\nu$ at the CERN
 SPS}  [CERN P-326], CERN-SPSC-2005-013, 
http://na48.web.cern.ch/NA48/NA48-3/ \\
Y.B.~Hsiung  {\it et al.}, 
{\em L.O.I. for the measurement of the $K_L \to \pi^0 \nu \bar\nu$ 
Branching Ratio at J-PARC}, 
http://www-ps.kek.jp/jhf-np/LOIlist/LOIlist.html


}
\end{thebibliography}
\end{document}